\documentclass[article,reprint,amsmath,amssymb,raggedbottom,letterpaper,final,superscriptaddress,citeautoscript,floatfix,notitlepage]{revtex4-2}

\usepackage{graphicx}
\usepackage{dcolumn}
\usepackage{bm}

\usepackage[usenames,dvipsnames]{color}
\usepackage{graphicx, microtype}
\usepackage[bookmarks=false,colorlinks]{hyperref}
\hypersetup{linkcolor=Plum,citecolor=RubineRed,filecolor=Plum,urlcolor=PineGreen}
\usepackage{float}
\usepackage{graphicx}
\usepackage{xspace}

\usepackage[usenames,dvipsnames]{color}
\usepackage{graphicx,microtype}

\usepackage[bookmarks=false,colorlinks]{hyperref}

\hypersetup{
    linkcolor=Plum,        
    citecolor=RubineRed,             
    filecolor=Plum,      	    
    urlcolor=PineGreen,         
}

\begin{document}

\title{Trigonal Symmetry Breaking and its Electronic Effects in Two-Dimensional Dihalides and Trihalides}

\author{Alexandru B.\ Georgescu}
\email{alexandru.georgescu@northwestern.edu}
\affiliation{%
Department of Materials Science and Engineering, 
Northwestern University, Evanston, Illinois 60208, USA}
\author{Andrew J.\ Millis}
\email{ajm2010@columbia.edu}
\affiliation{Center for Computational Quantum Physics, Flatiron Institute, 162 5th Avenue, New York, NY 10010, USA}
\affiliation{Department of Physics, Columbia University, 538 West 120th Street, New York, New York 10027, USA}%
\author{James M.\ Rondinelli}%
 \email{jrondinelli@northwestern.edu}
\affiliation{%
Department of Materials Science and Engineering, 
Northwestern University, Evanston, Illinois 60208, USA}

\date{\today}%

\begin{abstract}
We study the consequences of the approximately trigonal ($D_{3d}$) point symmetry of the transition metal (M) site in
two-dimensional van der Waals MX$_2$ dihalides and MX$_3$ trihalides. The trigonal symmetry leads to a 2-2-1 orbital splitting of the transition metal $d$ shell, which may be tuned by the interlayer distance, and changes in the ligand-ligand bond lengths. 
Orbital order coupled to various lower symmetry lattice modes may lift the remaining orbital degeneracies, and we explain how these may support 
unique electronic states using ZrI$_2$ and CuCl$_2$ as examples, and offer a brief overview of possible electronic configurations in this class of materials.
By building and analysing Wannier models adapted to the appropriate symmetry we examine how the interplay among trigonal symmetry, electronic correlation effects, and $p$-$d$ orbital charge transfer leads to insulating, orbitally polarized magnetic and/or orbital-selective Mott states.
Our work establishes a rigorous framework to understand, control, and tune 
the electronic states in low-dimensional correlated halides. 
Our analysis shows that trigonal symmetry and its breaking is a key feature of the 2D halides that needs to be accounted for in search of novel electronic states in materials ranging from CrI$_3$ to $\alpha$-RuCl$_3$.

\end{abstract}

\maketitle

\section{Introduction}

Transition metal compounds exhibit electronic properties of high scientific and technological interest, including ferroelectricity \citep{Schlom2007,Garcia2010,Karin2021,Alex2014}, quantum magnetism \citep{Tokura1998,Gibertini2019,DivineAlexNPJ,Molegraaf2009,Mizuguchi2008,AlexDivine20202}, metal-insulator transitions \citep{Landscapes,guzman,Shamblin2018,JMReview,Forst2015,Medarde1998, Georgescu2019,Claribel,VO2Length,Caviglia2013,Lauren2019} , and high transition-temperature superconductivity \citep{FisherEnergy,Lee2006,Norman2020,Gariglio2009,Keimer2015}. Transition metal oxides derived from the AMO$_3$ perovskite structure have been a focus of particular attention because any $3d$ or $4d$ transition metal can occupy the M site with (typically) partially filled $d$ shells, while variation of the A-site ion can tune the relative valence of the M site ion and the electronic bandwidth. The perovskite structure is also highly polymorphic.

It allows for many variants of the basic structure that exhibit different crystallographic symmetries, which activate interesting electronic states  and the pseudocubic structure allows for a wide variety of superlattices to be built \citep{VO2Length,Claribel}.
Basic to the electronic physics of perovskites is the cubic ($O_h$) point symmetry of the M-site ion and its reduction  to tetragonal symmetry by even parity octahedral distortions.

Recently, two-dimensional (2D) van der Waals transition metal MX$_2$ dihalides and trihalides MX$_3$ with X a halogen ligand have become of interest as they exhibit layer-dependent ferromagnetism as in VI$_3$ \citep{Huang2020,Son2019,Tian2019,Gati2019,VI3H,TuningMagnetism}, possible Kitaev spin liquid behavior in RuCl$_3$ \citep{Yadav2016,Yokoi2021,Suzuki2021,Li2021,Takagi2019,doi:10.1146/annurev-conmatphys-033117-053934} and other magnetic phenomena \citep{ReviewTheory2D,CrI3Stacking,VBr3}. Additionally, the crystal and electronic structures are highly two dimensional, so the materials can be exfoliated, made in monolayer form, doped by gating and layering with other compounds, and potentially twisted into Moir\'e materials \citep{Chen2020}. Similar to the perovskite transition metal oxides, the M site is six-fold coordinated by ligands and can host essentially all $3d$ and $4d$ transition metals often with partial $d$-orbital occupancy, which endows them with the aforementioned physical properties. Dihalide and trihalide compounds with the same transition metal will exhibit different nominal valence and octahedral coordination.  \autoref{tab:compounds} lists structural and electronic features of several known 2D halides.

Although the sixfold coordination of the metal cations in the dihalides and trihalides would suggest octahedral symmetry $O_h$, 
analogous to the octahedral coordination in many perovskite oxides, 
the halides are more appropriately described as trigonally coordinated.
A local orbital basis derived from $e_g$ and $t_{2g}$ representations of  the $O_h$ group is not the most useful description \citep{CrystalFieldBook}. 

As \autoref{tab:compounds} shows, the metal site symmetry in the bulk halides is either trigonal ($D_{3d}, D_3, C_{3i}$) or a trigonal subgroup ($C_{2h}$) or even lower \citep{2Dreview} and as previously noted for MX$_2$ compounds \citep{NormanMX2}  a basis that conforms to the symmetry enables a more straightforward treatment of the physics.

\begin{table}
\begin{ruledtabular}
\centering
\caption{\label{tab:compounds}
Symmetry, atomic structure, $d$-electron configuration for the metal, and magnetic order of experimental transition metal dihalides and trihalides. AFM, FM, and HM indicate antiferromagnetic, ferromagnetic, and helimagnetic, respectively. NA and MV indicate not applicable and  multiple values for the specified lengths, respectively. Structural data from Ref.\ \citep{ICSD} and magnetic data from Ref.\ \citep{2Dreview}.}
\begin{tabular}{llllll}
Structure   type    & CdI$_2$  & CdCl$_2$ & BiI$_3$   & CrCl$_3$ & AlCl$_3$           \\
\hline\\[-0.8em]
X  packing         & \emph{hcp}   & \emph{ccp}   & \emph{hcp}                   & \emph{ccp}   & $\sim$\emph{ccp}\\
Space   group       & $P\bar{3}m1$ & $R\bar{3}m$  & $R\bar{3}$ & $P3_212$ & $C2/m$          \\
Point   group       & $D_{3d}$   & D$_{3d}$   & $C_{3i}$    & $D_3$ & $C_{2h}$              \\[0.6em]

Compound            & FeI$_2$           & NiI$_2$          & FeCl$_3$     & CrCl$_3$ &     CrI$_3$                  \\
\hline\\[-0.8em]
M-$d^n$ config.\               & $d^6$    & $d^8$    & $d^5$                    & $d^3$    & $d^3$                    \\

M-site sym.\    & $D_{3d}$   & $D_{3d}$   & $C_3$ & $C_3$   & $C_2$ 	\\
Magnetism           & AFM      &    HM   &      HM                &    AFM   &       FM                \\
$L_1$                &  4.050  &   3.927  &      3.380 &     3.483          &     MV              \\
$L_2$                & 4.110   &   4.580  &     3.216      &   3.228         &    MV     \\
$L_3$                &  N/A     &     N/A     &     3.472      &    3.229   &     MV  \\[0.3em]
      
Compound            & TiCl$_2$ & MnCl$_2$ &          VCl$_3$    & $\alpha$-RuCl$_3$ &    $\alpha$-RuCl$_3$               \\
\hline\\[-0.8em]
M-$d^n$ config.\   & $d^2$    & $d^5$    & $d^2$                    & $d^5$    &         $d^5$           \\
M-site sym.\   & $D_{3d}$   &  $D_{3d}$    & $C_3$ & $C_3$   &  			$C_2$		\\
Magnetism           &  AFM     & AFM, HM       &     AFM        &  AFM     &      -                 \\
$L_1$             &   3.430  &   3.711    &    3.471    &   3.44    &          MV             \\
$L_2$             &  3.636 &  3.505     &   3.366     &   3.490    &         MV              \\
$L_3$             &   NA    &   NA   &        3.366   &   3.487    &             MV          \\
\end{tabular}
\end{ruledtabular}
\end{table}

In this paper we present a general analysis of the local electronic structure of the transition metal $d$ shells in dihalides and trihalides, considering both ideal structures and consequences of cooperative atomic displacements. By analogy to the more frequently  studied perovskite compounds, we identify what correlated electron behavior may arise. We focus on the implications of the trigonal point symmetry and answer the following questions: What is the appropriate orbital description for the correlated electrons? How to tune the orbital  structure  by varying  inter-layer distances and local MX$_6$ geometry, \emph{i.e.},  ligand-ligand and metal-ligand distances? What are the consequences of symmetry reductions for the correlated electron phases. 

In addressing these questions, it is often  helpful to construct a Wannier basis that includes a representation of the local transition metal $d$ orbitals, which host the electron interactions associated with correlated electron phenomena.  
However, standard Wannierization procedures for these materials,  do not easily produce a basis that transforms properly under trigonal symmetry operations. We present a procedure for constructing an appropriate symmetry-adapted basis, which can easily be adapted to other materials. 
This allows us to  discuss the correlated insulating states that may be obtained in stoichiometric compounds, as well as new electronic states that may appear as a result of further symmetry breaking, such as those present in the semi-1D chains in ZrI$_2$, and certain polymorphs of RuCl$_3$, CrI$_3$ and other halides \citep{2Dreview}.

\section{Local structure: electronic states and symmetry considerations \label{sec:symmetry}}   
\begin{figure}
\centering
\includegraphics[width=0.5\textwidth]{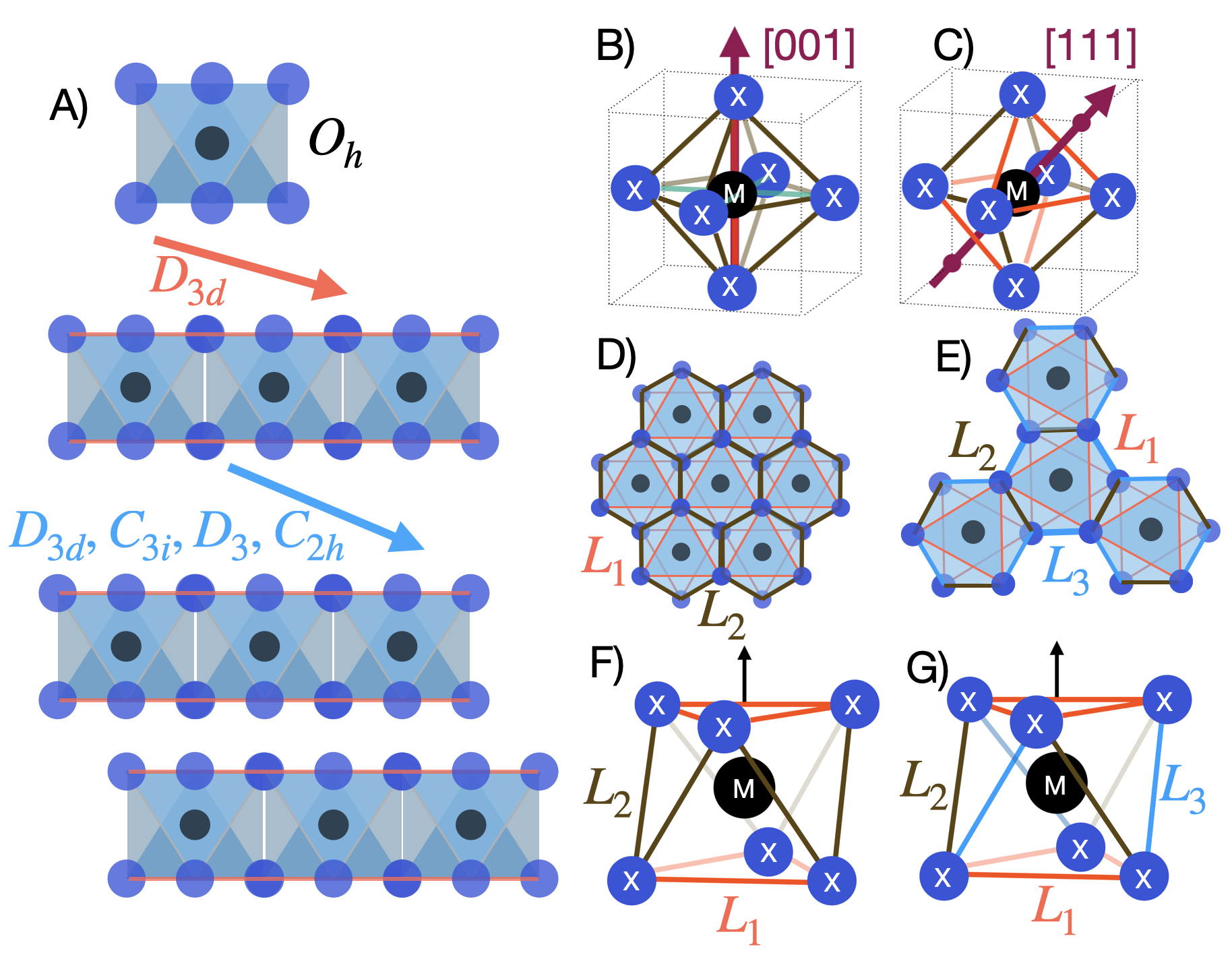}
\caption{A) Sequential illustration of symmetry breaking as a result of the layered structure in dihalides and trihalides. Placing an ideal octahedron  within a monolayer will reduce its O$_h$ symmetry to trigonal $D_{3d}$ symmetry: 3-fold rotations along the z axis are preserved, and so is mirror and inversion symmetry - even without distortions to the octahedron itself. 
Sketch of B) tetragonal and C) trigonal  symmetry breaking of an octahedron. Heavy magenta arrows indicate vector normal to monolayer planes.
Layer structure of MX$_6$ octahedra for D) MX$_2$ dihalides and E) MX$_3$ trihalides. Definitions for halide-halide lengths for octahedra in F) MX$_2$ and G) MX$_3$ halides. The red edges with length $L_1$ surround the face parallel to the monolayer planes, $L_2$ are edges that are not parallel to the material plane and that connect nearest-neighbor MX$_6$ octahedra. $L_3$ edges do not connect octahedra, are not parallel to the plane, and only appear in MX$_3$ compounds.
}
\label{fig:layering}
\end{figure}

The basic structural unit of 2D halides is a plane of transition metal ions, with each transition metal ion coordinated with 6 halogen ions (\autoref{fig:layering}A).
Unlike in the layered perovskite-based transition metal  compounds, the local  axes of the octahedron surrounding a transition metal ion in a halide are rotated with respect to the axes that define the two-dimensional plane. Specifically, choosing coordinates such that the monolayer plane is perpendicular to [001] then the octahedral axes (M-X bond directions) are 
[$\pm 1\pm 1\pm 1$] in the ideal halides
(compare panels B and C of \autoref{fig:layering}). In the ideal monolayer case, the point symmetry of the M ion is then $D_{3d}$.  
The symmetry may be further reduced by spin-orbit coupling, additional relaxations of the atomic positions, magnetic order and by different stackings of the layered structures. We discuss some lattice symmetry reductions, as well as magnetic order later. 
 
The key feature of the $D_{3d}$ symmetry is a three-fold rotation axis, in the ideal case perpendicular to the monolayer plane of MX$_6$ octahedra (\autoref{fig:layering}D,E) and passing through a triangular X$_3$ face  (\autoref{fig:layering}F,G).

In $D_{3d}$ symmetry, the M-X bond lengths  remain equal,  however, the X-X ligand distances $L_i$ ($i=1-3$) need not be equal. The inequivalence is  accommodated through changes in the X-M-X intra-octahedral angles. Similar effects are known in edge sharing perovskites \cite{Streltsov2017,PhysRevB.100.064101}
Because of the planar structure formed by edge connectivity of the octahedra, the halogen-halogen distances perpendicular to $(001)$ (denoted $L_1$ in panels F and G of \autoref{fig:layering}) are inequivalent to X-X distances on other triangular faces.

This can be plainly observed in the MX$_2$ compounds, where X-X edges not parallel to the plane connect different octahedra, while those that are parallel to the plane connect the same octahedra. This leads to the inequivalence of the $L_1$ and $L_2$  X-X edges (\autoref{fig:layering}F).

One may parametrize the trigonal distortion away from perfect octahedral coordination either by the $L_2/L_1$ ratio or by the X-M-X bond angles. We find that the  $L_2/L_1$ ratio is a more convenient parametrization in $D_{3d}$ symmetry, as it plays an analogous role to the $c/a$ ratio for the tetragonal symmetry 
and tunes the splitting between the doublet and singlet. 

For the MX$_3$ compounds only half of the X-X edges not parallel to the plane connect  MX$_6$ octahedra, leading to a total of three inequivalent edges, $L_1$, $L_2$, and $L_3$. \autoref{fig:layering}D illustrates that in both cases, there are only two types of inequivalent faces: those parallel to the planes of MX$_6$, with three $L_1$ edges, and those oblique to the plane with edges $L_1$, $L_2$ and  $L_3$.

While the symmetry of an ideal halide monolayer is $D_{3d}$,  different arrangements of the planes can lead to symmetry reductions to  $D_3$ or  $C_{3i}$ in the bulk compounds (see \autoref{tab:compounds}). These subgroups, however, lead to the  same trigonal orbital basis for the transition metal M $d$ orbitals. Further,  anisotropic semicovalent and van der Waals interactions or additional electronically driven orderings can lead to further symmetry reductions (see \autoref{tab:compounds}). In such cases the clearest way to understand the resulting orbital structure is as an additional symmetry reduction beyond trigonal.

\begin{table}
\begin{ruledtabular}
\centering
\caption{\label{tab:orbitals}
Relationships among the atomic $d$ orbitals for the standard cubic (tetragonal) basis with $z$ aligned to the (001) direction shown in \autoref{fig:layering}B and trigonal basis with $z$ axis aligned to (111) direction shown in \autoref{fig:layering}C. Note that in the absence of trigonal symmetry breaking, the $a_{1g}$ and e$_g^{\sigma}$ orbitals become degenerate and combine to transform as the $t_{2g}$ representation.
}
\begin{tabular}{llll}%
\multicolumn{2}{c}{Cubic Basis [001]} & \multicolumn{2}{c}{Trigonal Basis [111]} \\
\cline{1-2}\cline{3-4}

Symmetry	& Orbitals 	& Symmetry	& Orbitals \\
\cline{1-2}\cline{3-4}
$t_{2g}$	& $d_{\bar{x}\bar{y}}$	&	$a_{1g}$	& $d_{z^2}$ \\[0.3em]
		& $d_{\bar{y}\bar{z}}$	&	$e_g^\pi$			& $\frac{2}{\sqrt{6}}d_{xy} + \frac{2}{\sqrt{3}}d_{yz}$   \\[0.3em]
		& $d_{\bar{x}\bar{z}}$	&				& $\frac{2}{\sqrt{6}}d_{x^2-y^2} - \frac{1}{\sqrt{3}}d_{xz}$	\\[0.5em]
\cline{1-2}\cline{3-4}\\[-1em]
$e_g$	&	$d_{\bar{x}^2-\bar{y}^2}$		&	$e_g^\sigma$		& $\frac{1}{\sqrt{3}}d_{x^2-y^2} + \frac{2}{\sqrt6}d_{xz}$	\\[0.3em]	
& $d_{\bar{z}^2-r^2}$ & & $\frac{1}{\sqrt{3}}d_{xy} - \frac{2}{\sqrt6}d_{yz}$ \\
\end{tabular}
\end{ruledtabular}
\end{table}

The trigonal symmetry has implications for the electronic structure. Under $D_{3d}$  symmetry, the 5 $d$ orbitals transform as  two doublets ($ e_g^\sigma$ and $ e_g^\pi$) and a singlet ($a_{1g}$). The wave functions and level splittings are not constrained by symmetry and depend on details of atomic-scale physics. The trigonal structure of the halides may be viewed as a weak distortion of the $O_h$ symmetry familiar from cubic perovskites, enabling  a simpler interpretation of the basis functions.
Under $O_h$ symmetry, the atomic $d$ orbitals transform as an $e_g$ symmetry doublet (wavefunctions conventionally chosen as $d_{\bar{x}^2-\bar{y}^2}$ and $d_{3\bar{z}^2-r^2}$) and a $t_{2g}$ triplet (wavefunctions conventionally chosen as $d_{\bar{x}\bar{y}},d_{\bar{x}\bar{z}},d_{\bar{y}\bar{z}}$) where  $\bar{x},\bar{y},\bar{z}$ are the three octahedral axes. 
Ligand fields arising from hybridization lead to an energy separation between the doublet and triplet states of the order of 2\,eV. The further reduction of the symmetry from $O_h$ to $D_{3d}$ does not additionally split the $e_g$ states, which now form the $e_g^\sigma$ doublet representation  of $D_{3d}$.
It does, however,  split the $t_{2g}$ triplet into an $ e_g^\pi$ doublet  and an  $a_{1g}$ singlet with the $e_g^\pi-a_{1g}$ level splitting being typically smaller than the energetic separation to the $e_g^\sigma$ doublet. Then to first order in the trigonal distortion, the basis functions for the $e_g^\sigma$ representation are  linear combinations of the familiar cubic-basis $e_g$ states while the  basis functions for the $e_g^\pi$ and  $a_{1g}$ states are linear combinations of the familiar $t_{2g}$ states. 
(\autoref{tab:orbitals}) 
\footnote{Note that for the cubic case, the conventional coordinate system involves a $z$ axis  chosen parallel to an M-X bond while for the trigonal case the conventional coordinate system involves a $z$ axis passing through an octahedral face (\emph{i.e.}, along $[111]$ in conventional octahedral coordinates).
}.
Lowering the symmetry below trigonal, in particular by breaking the  $C_3$ rotational symmetry about the axis perpendicular to the plane, will lift the degeneracies of the two $e_g$ doublets. Such distortions may occur if the transition metal valence is such that one of the orbital pairs forming a doublet are partially filled, enabling an electronic symmetry-breaking transition.

\section{Physics of the  Trigonal Distortion in T\lowercase{i}C\lowercase{l}$_2$}

\begin{figure*}
    \centering
    \includegraphics[width=0.75\textwidth]{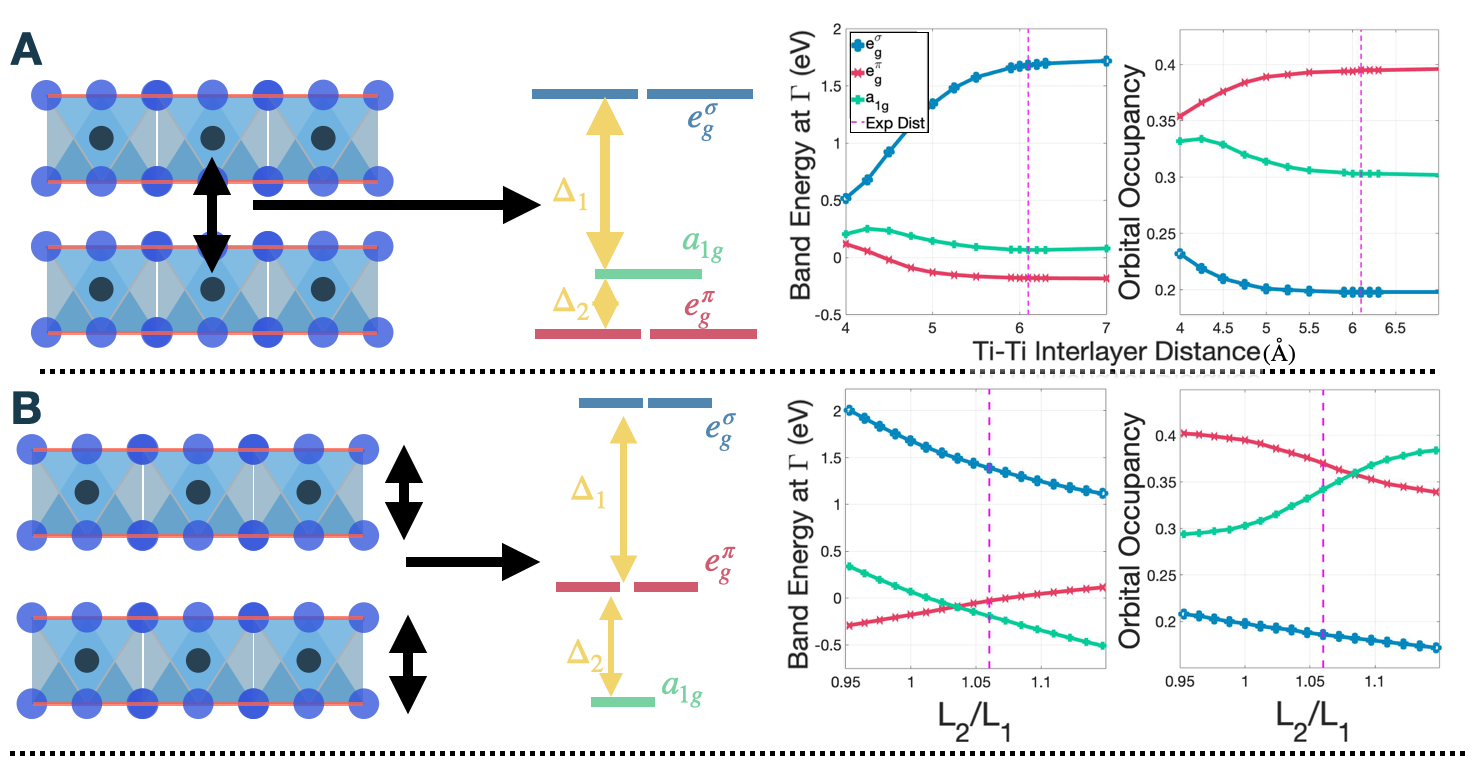}
    \caption{ Electronic orbital splitting at the non-magnetic DFT level for TiCl$_2$ as a function of (A) Ti-Ti interlayer distance and (B) $L_2/L_1$ ratio, as quantified by crystal field splitting at the $\Gamma$ point and orbital occupancy as obtained from the density matrix of the orbital projections. In A, the octahedron is kept perfect, with a X-X distance of 3.43\,\AA.  In B, the inter-layer distance is kept fixed to the experimental value, while $L_2/L_1$ is varied. The experimental inter-layer distance and $L_2/L_1$ ratio are indicated with broken magenta lines.
    }
    \label{fig:orbitalsplittingresult}
\end{figure*}

We now present density functional theory (DFT) calculations  on the representative dihalide TiCl$_2$ \footnote{DFT calculations were performed using the Quantum Espresso software package on a 3 atom formula unit, using ultrasoft pseudopotentials using the PBE exchange-correlation functional, an energy cutoff of 748eV and a 20x20x20 k-grid. Orbital occupations were obtained using the orthonormal atomic projections, as implemented in Quantum Espresso; DFT+U calculations were performed with the U applied to orthonormal atomic projections as well, within the Dudarev simplified +U formalism.}, using the experimental structure obtained from ICSD \citep{ICSD}.
TiCl$_2$ exhibits a $d^2$ electronic configuration and its primitive $P\bar{3}m1$ structure contains a single formula unit. The Ti atoms are aligned along $z$, making for the simplest possible stacking. The Ti site symmetry is $D_{3d}$. Details of its crystal structure are presented in \autoref{tab:compounds}. 

We begin by analyzing non-magnetic DFT calculations using the PBE functional, in order to isolate the effect of the atomic structure on the on-site crystal field splittings and orbital order before performing further analysis. We then investigate correlation effects in the DFT+U approximation. This stepwise approach allows us to disentangle the effects of the structure from that of $d-$shell electron-electron interactions.
We select  different interlayer distances, X-X distances, and $L_2/L_1$ ratios to probe the trigonal symmetry effects on the orbital structure. 
We characterize the electronic structure changes via the on-site energy of the orbitals at the $\Gamma$ point, and the orbital occupations, as obtained from the eigenvalues of the density matrix defined from orbital projectors.

Experimentally, these perturbations to the inter-layer distance may be realized via external pressure, while the $L_2/L_1$ ratio can be tuned via epitaxial strain.  
The effects we find pertain to MX$_3$ compounds as well, as the MX$_3$ structure can be obtained from the MX$_2$ structure by removing 1/3 of the M atoms and keeping the ligand octahedral structure intact. 

\subsection{Level Splittings and Orbital Occupancies}
 
\autoref{fig:orbitalsplittingresult} presents a summary of our results.

Starting with the experimental structure, we find that the energy splitting at $\Gamma$  corresponds to the expected 2-2-1 orbital  splitting as shown by the vertical broken line in \autoref{fig:orbitalsplittingresult}A,B.

Upon visualizing isosurfaces of the wavefunctions in real space at the $\Gamma$ point (not shown), we find the orbitals represent wavefunctions of the trigonal basis.

These wavefunctions are similar to the Wannier functions presented in \autoref{fig:DOS}.

The lowest energy Ti $d$-derived state at  $\Gamma$ is the $a_{1g}$ state. The wavefunction exhibits lobes directed through the faces parallel to the plane. The next two higher energy states are the  $e_g^\pi$ doublet with wavefunction lobes directed at the faces of the octahedron transverse  to the plane. The highest energy states transform as the $e_g^\sigma$ doublet in which the orbital lobes are directed along the octahedral axes, pointing towards the Cl anions.
This behavior is expected because they behave equivalently to the $e_g$ states of the cubic basis. The $e_g^\sigma$-$e_g^\pi$ splitting of approximately $1.5$\,eV is somewhat less than the approximately $2-3$ eV splitting typical in perovskite transition metal oxides, reflecting the larger energy separation and weaker hybridization of the metal $d$ states with the halogen ligand $p-$states.

Next, we examine the effects on the electronic structure of varying the structural parameters (\autoref{fig:orbitalsplittingresult}). We begin by fixing the TiCl$_6$ octahedra to  $L_1=L_2= 3.430$\,\AA\ with all M-X bonds fixed to 2.425\,\AA\ and vary the inter-layer distance. We find that the  crystal-field splitting between the $e_g^\sigma$ and the $e_g^\pi/a_{1g}$ complex increases with increased inter-layer distance as does the corresponding  difference in orbital occupancy (\autoref{fig:orbitalsplittingresult}A). The $e_g^\pi-a_{1g}$ splitting is less dependent on the interlayer distance. For inter-layer distances greater than or equal to the experimental value measured at ambient pressure, the crystal field splitting and orbital occupations converge to their asymptotic isolated layer values. 

Next, we keep the inter-layer distance fixed to the ambient pressure equilibrium experimental value of 6.1\,\AA\ and vary the $L_2/L_1$ ratio by displacing the Cl atoms along $z$.
This changes $L_2$ while keeping $L_1$ the same. 
We find that increasing $L_2$ increases the energy difference between the highest two and lowest three orbital states states. 
In contrast to the effect of increasing the inter-layer distance, changing $L_2/L_1$ also decreases the $a_{1g}$ orbital energy relative to the $e_{g}^\pi$ energy. 
The occupancies change in the corresponding manner. The relative sizes of the energies and the orbital splittings, however, are not fixed.  
Importantly, we find that the critical $L_2/L_1$ ratio at which the occupations are equal occurs at a different critical $L_2/L_1$ ratio that gives a vanishing energy difference for $e_g^\pi-a_{1g}$ orbitals. 
Both correspond to $L_2/L_1 \neq 0$, underlining that the symmetry of the representation of the Ti $d$-shell is always at most trigonal, and never cubic.

\subsection{Orbital Physics in the Trigonal Wannier Basis}

Key to understanding the physical effects in a correlated a material is an appropriate tight-binding model, which can be constructed via a Wannierization procedure as implemented in standard electronic structure codes. A Wannierization is in effect a choice of a basis that represents band states in a certain energy range, along with a projection of the density functional Hamiltonian onto this basis. It is desirable to choose a basis that is adapted to the physics of the problem at hand; both because an appropriate choice provides physical insight and because the form and magnitude of beyond-DFT interactions depends on the basis chosen.

Standard application of the Wannier90 \citep{wannier90} code to di- and trihalide compounds tends to lead to the $e_g$-$t_{2g}$ basis functions well adapted to $O_h$ symmetry. To obtain Wannier functions adapted to trigonal symmetry, we find it is best to first use Wannier90 to obtain a basis, and then rotate the basis to obtain a diagonal density matrix. Another option is to diagonalize the Wannier Hamiltonian at the $\Gamma$-point. A third choice is to diagonalize the on-site term in the real-space Wannier Hamiltonian.  To obtain real-space isosurfaces, we used the eigenvectors obtained by diagonalizing the density matrix to build a linear superposition of the Wannier functions obtained directly from Wannier90.

\begin{figure}
    \centering
    \includegraphics[width=0.49\textwidth]{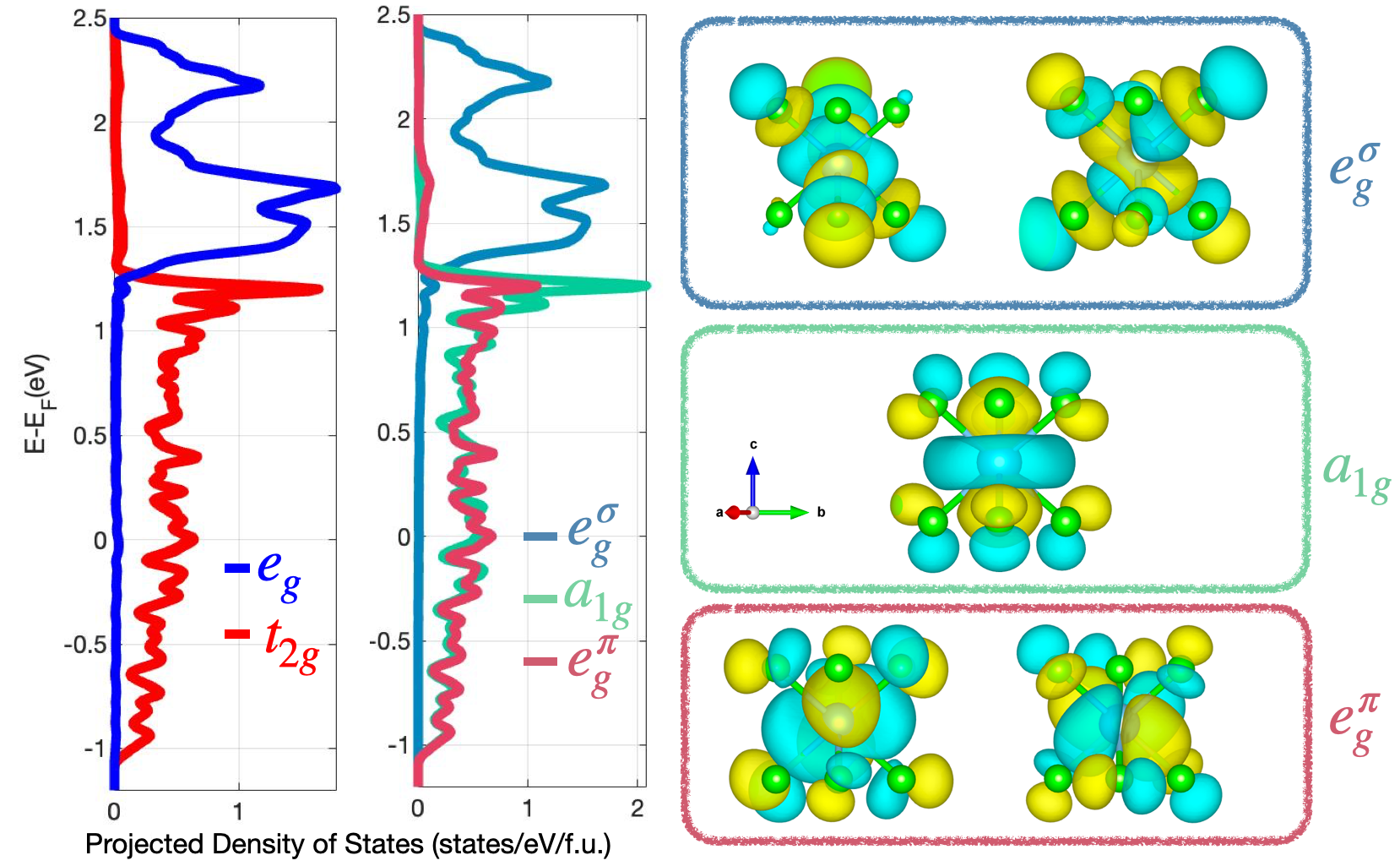}\\
    \caption{Projected density of states for the Wannier functions for TiCl$_2$ in the tetragonal basis (left) and trigonal basis (center).
    Eigenstates of the density matrix, as obtained by diagonalizing the density matrix (right).
    }
    \label{fig:DOS}
\end{figure}

\autoref{fig:DOS} shows the density of states projected onto different combinations of the Wannier states obtained for TiCl$_2$ in a standard application of Wannier90 (left panel) and in the basis that diagonalizes the density matrix (center panel) after the rotation. The total density of states is the same in both cases. The  $e_g^\sigma$-derived states, which are derived from the same functions in both cases, are approximately the same. The trigonal functions allow us to distinguish  the higher-lying $a_{1g}$ states from the lower-lying $e_g^\pi$ states.  \autoref{fig:DOS} also shows the corresponding isosurfaces of the resulting Wannier functions, which we obtain through linear mixing.

 \begin{figure}
    \centering
    \includegraphics[width=0.45\textwidth]{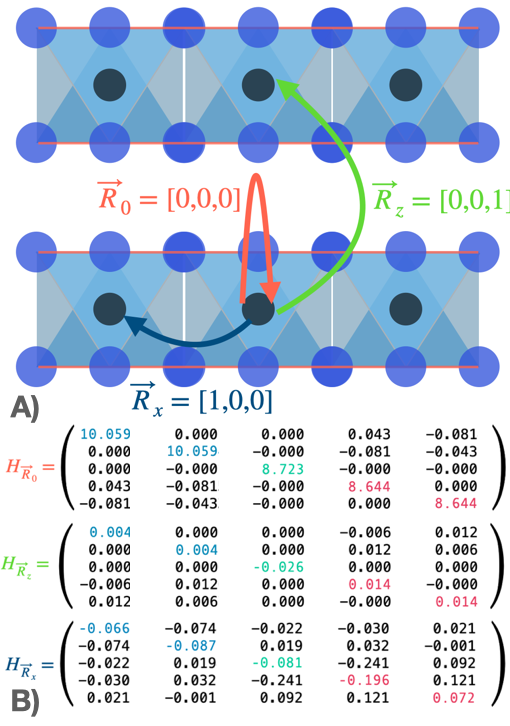}
    \caption{ A) $\vec{R}$ connecting  transition metals M, for which we show the hopping matrices of the Hamiltonian below in B). The diagonal elements are shown in blue for $e_g^\sigma$, a$_{1g}$ in green, and $e_g^\pi$ red.
    }
    \label{fig:hamiltonian}
\end{figure}

We now analyze the resulting Hamiltonian $H_{\vec{R}_i}$ (\autoref{fig:hamiltonian}), where the rows (columns) correspond to the following $d$ orbitals: the first two correspond to $e_g^\sigma$, the middle one to $a_{1g}$ and the last two to $e_g^\pi$. 
For $\vec{R}=(0,0,0)$, we find that the $a_{1g}$ orbital is orthogonal to the other sets of orbitals in $H_{\vec{R}_0}$, similar to what is obtained in the atomic orbital projection matrix. While within the pairs of degenerate doublets the mixing is practically 0, there is however mixing between the two pairs. This $H_{\vec{R}_0}$ however possesses the 2-2-1 pattern of eigenvalues consistent with D$_{3d}$ symmetry.  Hopping along $z$ between the Ti atoms $\vec{R}=(0,0,c)$ is primarily driven by $a_{1g}$-$a_{1g}$ hopping, hopping between the two pairs of doublet states, as well as small but non-zero hopping between equivalent orbitals as gleaned from $H_{\vec{R}_z}$ in \autoref{fig:hamiltonian}B. 
Finally, hopping in-plane between the atoms $\vec{R}=(a,0,0)$ is driven by non-zero hoppings between all orbitals, however the lowest three orbitals in $H_{\vec{R}_x}$ are the largest, as expected from the spatial orientation of the orbital lobes discussed previously. We also note that one of the eigenvalues of the hopping Hamiltonian is close to 0. The zero eigenvalue pertains to hopping among the highest two energy ($e_{g}^\sigma$) orbitals. As a result, there will always be a weakly dispersing flat band among the higher $e_g$ states along high symmetry directions in the zone.

\subsection{Lower structural symmetries}

Some of the dihalides and trihalides exhibit symmetries lower than trigonal. The point group $C_{2h}$, a subgroup of the trigonal point group $D_{3d}$, appears in both MX$_2$ and MX$_3$ compounds, namely in ZrI$_2$, as well as in some polymorphs of RuCl$_3$ and CrI$_3$. This symmetry reduction includes a breaking of the rotational symmetry, which splits the doublets, and may lead to minor mixing of the orbital eigenvalues; generally, subgroups of $D_{3d}$ which do not maintain the 3-fold rotational symmetry lead to a breaking of the doublets' degeneracy.   

 \begin{figure}
    \centering
    \includegraphics[width=0.49\textwidth]{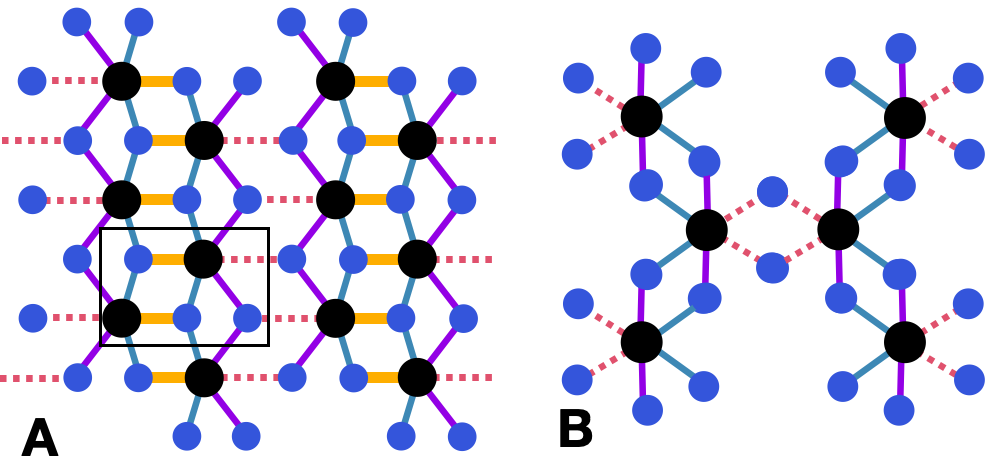}
    \caption{Single layer lattice structure for materials with point group symmetry C$_{2h}$, which breaks the degeneracy of the doublets: A) a layer of MX$_2$, with the structure from a layer of bulk ZrI$_2$, black rectangle represents the new unit cell B) a layer of MX$_3$, with a similar pattern representing bulk layers, for example  RuCl$_3$ and CrI$_3$ with the AlCl$_3$ structure type; all atoms in the sketch are within one unit cell. Distances between the atoms are indicated as follows: the red dotted lines are the longest, followed by purple, blue, then mustard yellow. The distortions characteristic of ZrI$_2$ are much stronger than those of the MX$_3$ materials. The M site symmetries are $C_{1h}$ for panel A and $C_2$ for panel B. No d-orbital degeneracies are enforced by either site symmetry.}
    \label{fig:C2h}
\end{figure}

\autoref{fig:C2h} shows that the single layers of ZrI$_2$ and RuCl$_3$ or CrI$_3$ no longer have three-fold rotational symmetry. The in-plane lattice parameters defining the formula unit of a di- or tri-halide lattice are no longer equivalent, and are replaced by perpendicular vectors describing the supercell that accomodates for this lower symmetry as described in the caption for the figure. This breaks the symmetry of the remaining two orbital doublet pairs. This symmetry reduction is associated with inequivalence of both the X-X bonds and the M-X bonds as indicated in \autoref{fig:C2h}. In ZrI$_2$ this is plainly discerned as the bond distortions are sufficiently strong that the structures can be understood to form 1D zig-zag chains \citep{ZrI2Chain}. A minor mixing between the $a_{1g}$ orbital and the other orbitals appears as a result of the symmetry lowering, and performing a Wannierization shows us that this minor mixing breaks its rotational symmetry.

The ZrI$_2$ structure also has additional forms of symmetry breaking beyond that leading to the $C_{2h}$ point group; for example, the two layers that form the minimum structural unit are not completely equivalent. The octahedral sizes differ between the layers. The symmetry of ZrI$_2$ is even lower in certain experimentally reported structures with two non-equivalent octahedra forming each layer leading to point group $C_{2v}$.  Such further symmetry reductions are beyond the scope of this paper.

\section{Electronic Correlations and Local Structure Effects} 
\begin{figure}
    \centering
    \includegraphics[width=0.49\textwidth]{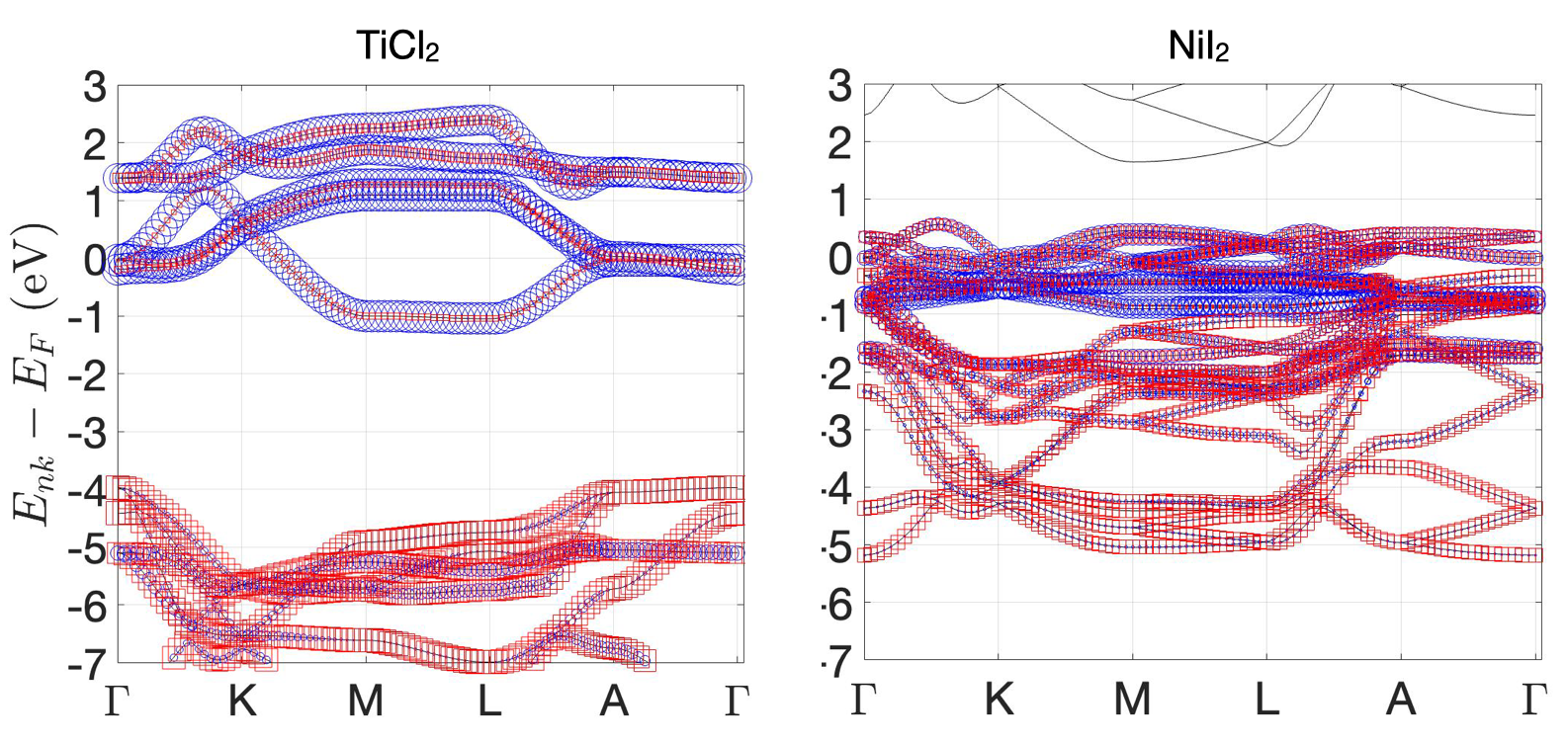}
    \caption{DFT Band structure and Wannier `fat bands' for a full $p-d$ model for  (left) TiCl$_2$ and (right) NiI$_2$, exemplifying materials that are in the Mott-Hubbard (TiCl$_2$) and charge-transfer (NiI$_2$) regimes. Blue circles correspond to transition metal M $d$ state projections, red squares to ligand $p$ states. }
    \label{fig:band-structure}
\end{figure}

Many transition metal compounds exhibit correlated insulating states. The correlation physics that produces these states tends to favor high spin, filled and empty state configurations; the exact order is determined by a combination of correlation effects and electron-lattice coupling. 
Similar to perovskite oxides, 2D di- and tri-halides can be in the Mott-Hubbard or in the charge-transfer regime of the Zaanen–Sawatzky–Allen (ZSA) classification \citep{Zaanen1985}. TiCl$_2$ is a clear example of a Mott-Hubbard material as the $d$ and $p$ states are clearly separated by ~4 eV (\autoref{fig:band-structure}A). Similar to the transition metal perovskite case, going right along the periodic table, \emph{i.e.}, towards higher orbital filling, the $p$-$d$ splitting decreases. In \autoref{fig:band-structure}B, we find the $p$-$d$ splitting is low for the well-known material NiI$_2$ \citep{Ronda1987,Physics1986,VanDerLaan1986} with the $p$ and $d$ bands overlapping in energy, corresponding to this insulating material being in the charge transfer regime. 

 \begin{figure}
    \centering
    \includegraphics[width=0.49\textwidth]{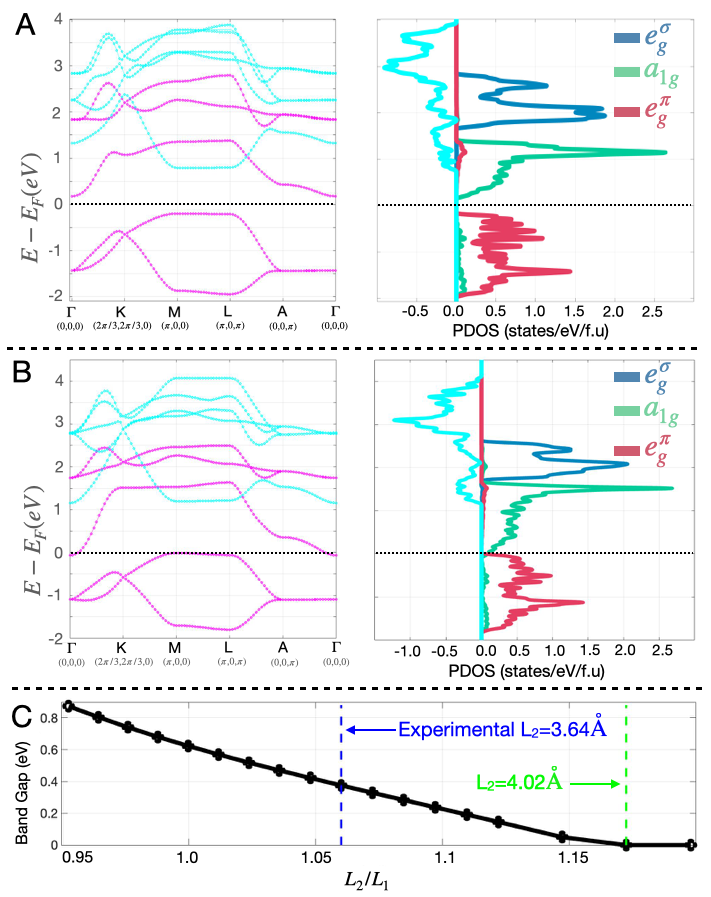}\vspace{-6pt}
    \caption{Electronic band dispersions and orbital projected density of states for ferromagnetic, high-spin TiCl$_2$ at the DFT$+U=2$\,eV level: (A) experimental structure and (B) structure with $L_2$ extended to 4.02\,\AA and inter-layer distance kept constant. Left: Bands, majority spin in magenta, minority spin in cyan. Right: projected density of states (PDOS), minority spin rescaled by dividing by 5. Increasing $L_2/L_1$ closes the gap by reducing the orbital polarization of the lower three orbitals. (C) Dependence of the electronic band gap in TiCl$_2$ with $L_2/L_1$ ratio for a fixed inter-layer distance. }
    \label{fig:U2FM}
\end{figure}

\subsection{Example of TiCl$_2$}

 We performed $\mathrm{DFT}+U=2$\,eV calculations on the dihalide TiCl$_2$ in the experimental structure and allowing for spin symmetry breaking to gain initial insight into the role of correlations. The  Hubbard on-site Coulomb interaction allows the opening of a gap in the electronic spectrum ($\approx$\,0.375\,eV), creating a fully orbital and spin polarized state (\autoref{fig:U2FM} A).
Taking the TiCl$_2$ experimental structure and allowing for spin-symmetry breaking, we find a high spin orbitally ordered state.
The minority spin channel is completely unoccupied. Two electrons reside in the majority spin channel of the two $e_{g}^\pi$ orbitals, while the $e_{g}^\sigma$ and $a_{1g}$ occupations are 
essentially zero. 

Calculations with structures obtained by elongating $L_2$ while keeping $L_1$ constant closes this gap, while shortening $L_2$ further opens the gap, by shifting the relative energy levels of the lowest three states (\autoref{fig:U2FM}). Similar effects can be obtained by changing the inter-layer distance as well. Importantly, we note that in the presence of electronic correlations, the electronic configuration favors the state that leads to fully occupied and empty states and the opening of a band gap, and a strong trigonal distortion is needed to counteract this effect. We note that, in a tetragonal basis, as discussed before, the lower three $t_{2g}$ states would remain equivalent, as their lobes would point along equivalent directions, which would obscure the identity of the active orbitals -- the singlet and doublet -- participating in the band gap opening. 

\subsection{Survey of Broken Symmetry Phases}
\begin{figure*}
    \centering
    \includegraphics[width=0.75\textwidth]{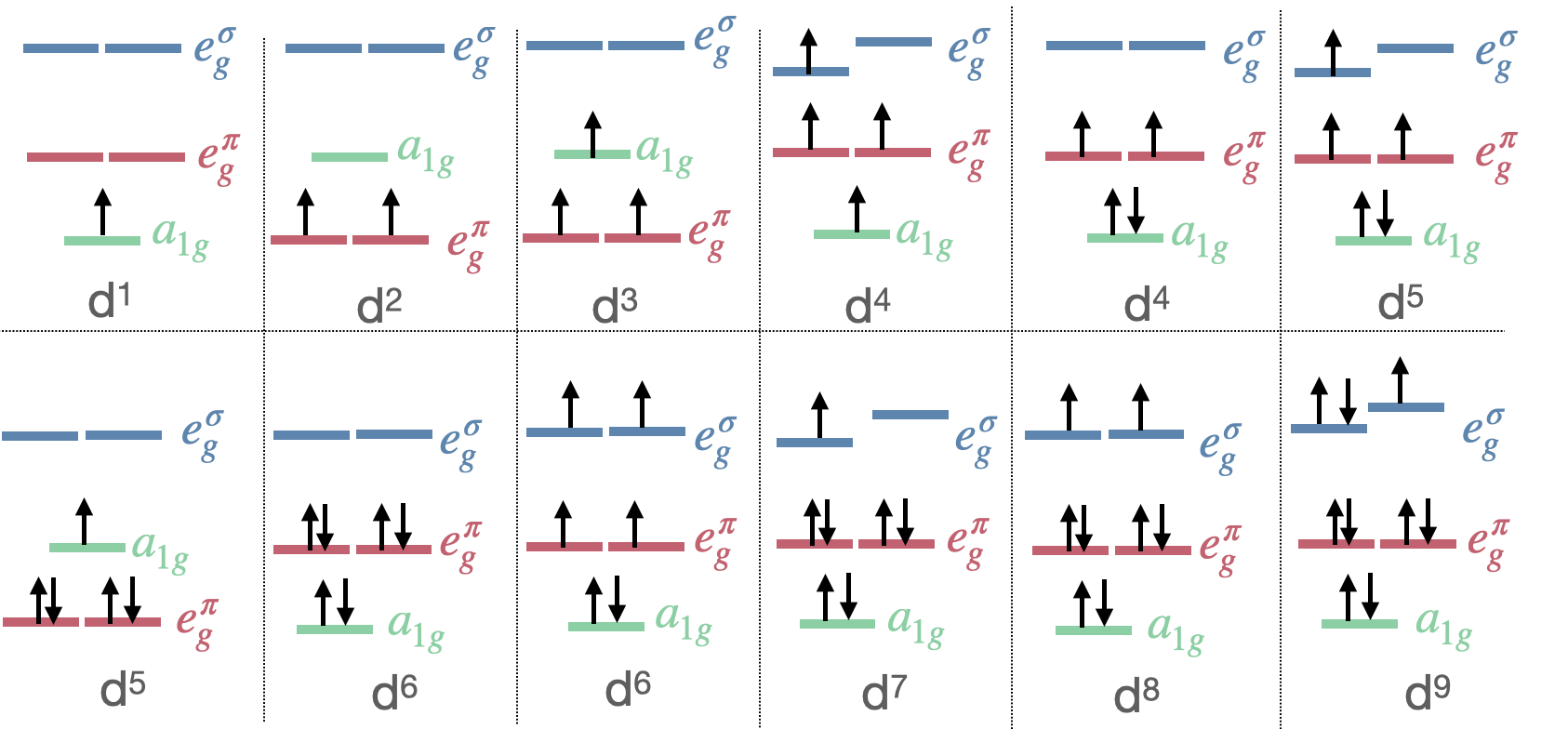}
    \caption{Different possible electronic states for $d^1-d^9$ occupations as discussed in the main text. Configurations that break the doublet degeneracy would be associated with further lattice distortions, reducing symmetry below trigonal. States with higher $a_{1g}/e_g^\pi$ occupation ratios are likely to have a higher $L_2/L_1$ ligand bond ratio.}
    \label{fig:electronconfig}
\end{figure*}

By varying the transition  metal ion and considering both dihalides and trihalides, one can obtain all $d^n$ configurations ranging from $d^1$ (TiI$_3$) to $d^9$ (CuCl$_2$, proposed below). 
As a result, a range of  correlated insulating states  can be produced, as shown in \autoref{fig:electronconfig}. Some of these states will involve orbital order that breaks the trigonal symmetry, leading to the appearance of one dimensional lattice structures - or, more generally, structures that break the three-fold rotational symmetry.

In the $d^1$ configuration the natural Mott insulating  state involves  an electron in the $a_{1g}$ orbital. This state preserves the trigonal symmetry. However the multiple superexchange pathways involving also the $e_g^\pi$ orbitals suggest that unless $L_2/L_1$ can be made very small the ground state will be ferromagnetic. However if the $L_2/L_1$ ratio can be made large enough, a change in level ordering may occur and it is possible to have an  orbitally ordered Mott insulator with one electron in the $e_g^\pi$ orbitals and a corresponding orbital order and trigonal symmetry breaking lattice distortion. 

In  the $d^2$ case, we may generically expect a high spin  ground state with two electrons in the $e_g^\pi$ oribitals, as found  in the DFT$+U$ calculations for TiCl$_2$. For a small enough $L_2/L_1$ ratio, however,  a state with one electron in the $a_{1g}$ orbital and one in $e_g^\pi$ doublet may lift the trigonal symmetry.

The relatively smaller 
ligand field splitting in the halides relative to the octahedral perovskite oxides suggests that the  $d^4,~d^5, d^6$  state will all be high spin. The $d^4$ and $d^6$ configurations would also exhibit orbitally ordered and trigonally broken states; $d^6$ can also support an insulating diamagnetic state with the gap opened by the crystal field splitting between the 2 higher energy and lower 3 orbitals. 

The $d^7$ state is most likely to be a $S=1/2$ state with the lowest three orbitals fully occupied, and one of the $e_g^\sigma$ orbitals half-filled and spin-polarized, likely leading to a structurally broken state of the form found in the materials with $C_{2h}$ point symmetry. The $d^8$ configuration is most likely to be a $S=1$ state with each $e_g^\sigma$ orbital half-filled and spin-polarized. In some scenarios, it may be orbitally-polarized and non-magnetic, associated with a symmetry-breaking mode in the lattice, most likely as previously discussed. 

Such configurations may be susceptible to metal-insulator transitions (MIT), which we assess using a recently devised machine-learning classification model \citep{AlexMLMIT}.
We found that the binary MIT-non MIT classifier tends to predict most of the 2D halides are candidate MIT compounds, giving a positive MIT classification for CrCl$_3$, FeCl$_3$, IrBr$_3$, MnCl$_2$, RuCl$_3$, TiCl$_2$, VCl$_3$ and ZrI$_2$, and a negative classification for CrI$_3$, FeI$_2$ and NiI$_2$. This is likely due to the similarity of this class of materials to perovskite oxide MIT transition compounds, both in the structural features exhibited, such as the average transition metal-ligand distances which range between 2.4\,\AA\ and 3.01\,\AA, the metal-metal distances of 3.3-4.2\,\AA, and electronic descriptors, e.g., estimated unscreened Hubbard $U$ values. 
However, due to their intrinsic broken symmetry and reduced dimensionality, these materials generally will be exclusively insulating. Similar to MIT compounds, the 2D halides will tend to display coupled electronic and lattice transitions at low temperatures \citep{Gati2019}.  

 \begin{figure}
    \centering
    \includegraphics[width=0.49\textwidth]{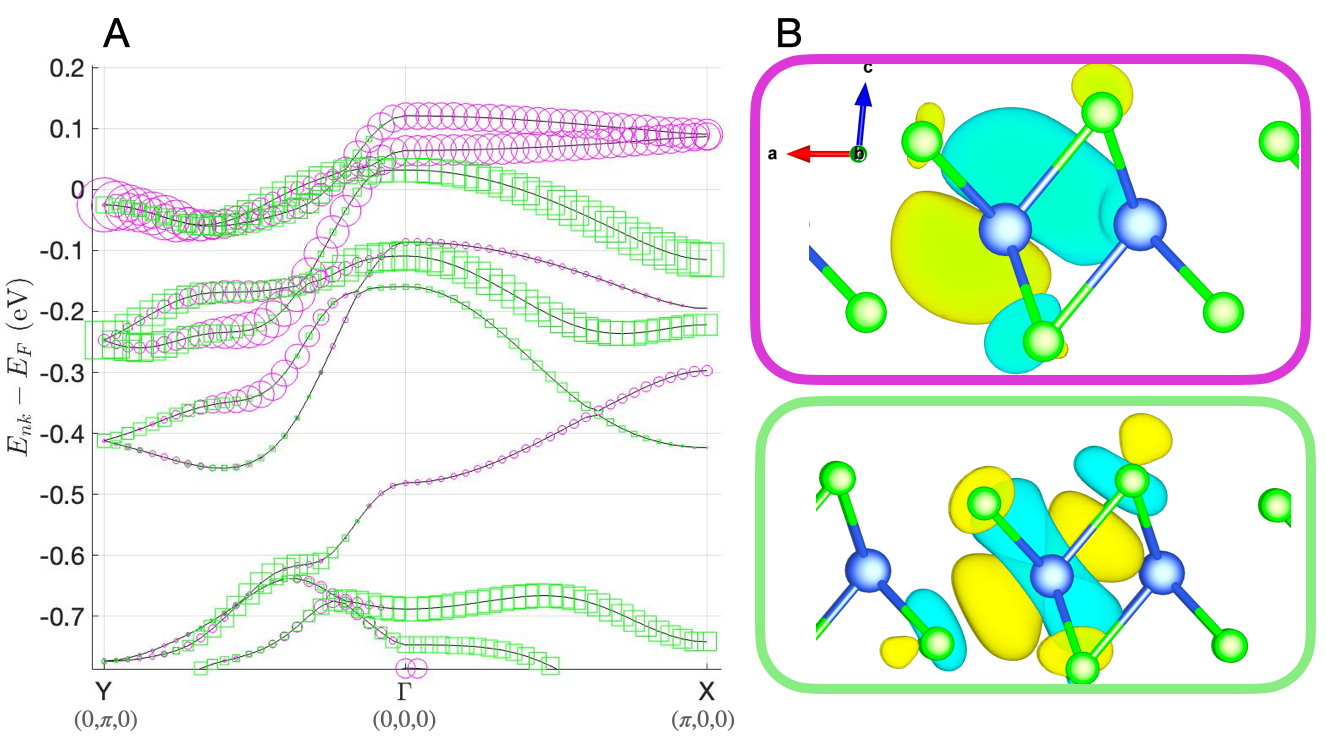}
    \caption{A) DFT band structure, and 'fat bands' of the top two Wannier orbitals on the hypothetical CuCl$_2$ monolayer discussed in the main text. B) Real-space isosurfaces of the highest energy d-orbitals, with occupations 0.912 and 0.93 per spin channel. Due to the pseudo-1D nature of the symmetry broken lattice structure, there is relatively little dispersion along the $\Gamma$-X direction.}
    \label{fig:CuCl2}
\end{figure}

For the $d^9$ and possibly $d^4$ configurations, 
the presence of one hole in the $e_g^\sigma$ manifold implies a breaking of rotational symmetry and likely, a one dimensional electronic structure. 
To examine this possibility, we perform the following simulation. We take a single layer of the experimental ZrI$_2$ structure
and replace Zr with Cu and I with Cl without allowing the structure to relax. Owing to the strong charge-transfer character of the resulting theoretical material, the Cu $d$ manifold
is mostly filled. Nonetheless, after performing the appropriate rotation of the $d$-shell basis, we find that the two least-occupied orbitals are $e_g^\sigma$ like (\autoref{fig:CuCl2}). These orbitals display non-zero orbital polarization with one orbital with $d_{x^2-y^2}$ character pointing along Cu-Cl bonds along the 1D chain, and one orbital of $d_{3z^2-r^2}$ character pointing towards Cu-Cl bonds that connect the 1D chains. This type of symmetry breaking can likely be exploited, provided the appropriate material strains, to form 1D conducting chains out of a 2D structure. This behavior is similar to how the pseudo-2D electronic structure of cuprate-like high-temperature superconductors forms from out of 3D conducting building blocks. Another orbitally polarized state can also be obtained with a different orbital basis for the $d^9$ configuration \citep{Qin2021}.

\section{Conclusions}
We showed that the highest possible metal orbital symmetry of 2D dihalides and trihalides comprising edge-shared MX$_6$ octahedra is trigonal, and analysed the interplay of atomic lattice, orbital physics and correlation effects. 
Within this trigonal basis, we showed that the amplitude of the electronic orbital splittings can be tuned by both the interlayer distance, as well as through changing the ligand-ligand bond ratio $L_2/L_1$, and that the effect of correlations strongly favors ordered states.

In addition, the orbital occupancies can be sensitively tuned
through changes in these atomic structure features, 
which can be achieved experimentally via pressure, strain, as well as possibly via optical excitations of the relevant structural modes to enable control of magnetic configurations and other electronic ordering. We showed that these materials can be analyzed from the point of view of the ZSA classification. 

We showed how one can build and analyze a Wannier model corresponding to this reduced symmetry, symmetry which can lead to novel (correlated) electronic states.
Our work serves as a basis to understand correlated phenomena in this class of materials, and their interplay with lattice symmetry modes, and easily allows for models to disentangle their roles - similar to recently built models on perovskites and their Ruddlesden-Popper phases \citep{Landscapes, Georgescu2019,Oleg,Han2018}. Our results may be particularly important in the search for candidate spin-liquis. Spin liquid states are predicted to be found in the vicinity of a Mott transition and in symmetric highly-frustrated structures. Our generic finding of strong orbital ordering and associated trigonal symmetry breaking will be important in appropriately modeling - and discovering - potential spin liquids and other novel states.

\begin{acknowledgments}
This research was supported in part by the National Science Foundation (NSF) under DMREF Award DMR-1729303.
The information, data, or work presented herein was funded in part by the Advanced Research Projects Agency-Energy (ARPA-E), U.S.\ Department of Energy, under Award Number DE-AR0001209.AJM is  supported in  part by Programmable Quantum Materials, an Energy Frontier Research Center funded by the U.S. Department of Energy (DOE), Office of Science, Basic Energy Sciences (BES), under award DE-SC0019443. 
The views and opinions of authors expressed herein do not necessarily state or reflect those of the United States Government or any agency thereof. The Flatiron Institute is a division of the Simons Foundation.

\end{acknowledgments}

\bibliography{Mendeley,more_refs}

\end{document}